 \documentclass[sigconf]{acmart}

\AtBeginDocument{%
  \providecommand\BibTeX{{%
    \normalfont B\kern-0.5em{\scshape i\kern-0.25em b}\kern-0.8em\TeX}}}

\copyrightyear{2023} 
\acmYear{2023} 
\setcopyright{acmlicensed}\acmConference[ICMI '23]{INTERNATIONAL CONFERENCE ON MULTIMODAL INTERACTION}{October 9--13, 2023}{Paris, France}
\acmBooktitle{INTERNATIONAL CONFERENCE ON MULTIMODAL INTERACTION (ICMI '23), October 9--13, 2023, Paris, France}
\acmPrice{15.00}
\acmDOI{10.1145/3577190.3614133}
\acmISBN{979-8-4007-0055-2/23/10}

\usepackage{amsmath}
\usepackage{subcaption}
\usepackage{enumitem}

\begin{document}



\settopmatter{printacmref=false}
\setcopyright{none}
\renewcommand\footnotetextcopyrightpermission[1]{}
\pagestyle{plain}
\title{ReNeLiB: Real-time Neural Listening Behavior Generation for Socially Interactive Agents}







\author{Daksitha Withanage Don}
\orcid{0009-0003-8831-9795}
\affiliation{
\institution{University of Augsburg,}
\institution{German Research Center for Artificial Intelligence}
\streetaddress{Universitätsstraße 6a}
\city{Augsburg}
\country{Germany}
\postcode{86159}, 
 }
 \email{daksitha.withanage.don@uni-a.de}

 \author{Philipp Müller}
\affiliation{
  \institution{German Research Center for Artificial Intelligence}
  \streetaddress{Saarland Informatics Campus}
   \city{Saarbruecken}
   \country{Germany}}
   \postcode{66123}
 \email{philipp.mueller@dfki.de}

 \author{Fabrizio Nunnari}
\affiliation{
  \institution{German Research Center for Artificial Intelligence}
  \streetaddress{Saarland Informatics Campus}
   \city{Saarbruecken }
   \postcode{66123}
   \country{Germany}}
   
 \email{fabrizio.nunnari@dfki.de}

 \author{Elisabeth André}
\affiliation{
  \institution{University of Augsburg}
  \streetaddress{	Universitätsstrasse 6a}
   \city{Augsburg}
   \postcode{86159}
   \country{Germany}}
 \email{andre@informatik.uni-augsburg.de}

\author{Patrick Gebhard}
\affiliation{
  \institution{German Research Center for Artificial Intelligence}
  \streetaddress{Saarland Informatics Campus}
   \city{Saarbruecken}
   \postcode{66123}
   \country{Germany}}
 \email{patrick.gebhard@dfki.de}







\renewcommand{\shortauthors}{Anonymous, et al.}

\begin{abstract}

Flexible and natural nonverbal reactions to human behavior remain a challenge for socially interactive agents (SIAs) that are predominantly animated using hand-crafted rules. 
While recently proposed machine learning based approaches to conversational behavior generation are a promising way to address this challenge, they have not yet been employed in SIAs. 
The primary reason for this is the lack of a software toolkit integrating such approaches with SIA frameworks that conforms to the challenging real-time requirements of human-agent interaction scenarios. 
In our work, we for the first time present such a toolkit consisting of three main components: (1) real-time feature extraction capturing multi-modal social cues from the user; (2) behavior generation based on a recent state-of-the-art neural network approach; (3) visualization of the generated behavior supporting both FLAME-based and Apple ARKit-based interactive agents.
We comprehensively evaluate the real-time performance of the whole framework and its components.
In addition, we introduce pre-trained behavioral generation models derived from psychotherapy sessions for domain-specific listening behaviors. Our software toolkit, pivotal for deploying and assessing SIAs' listening behavior in real-time, is publicly available. Resources, including code, behavioural multi-modal features extracted from therapeutic interactions, are hosted at \url{https://daksitha.github.io/ReNeLib}

\end{abstract}



\begin{teaserfigure}
  \centering
  \begin{subfigure}[b]{0.3\linewidth}
    \includegraphics[width=\linewidth]{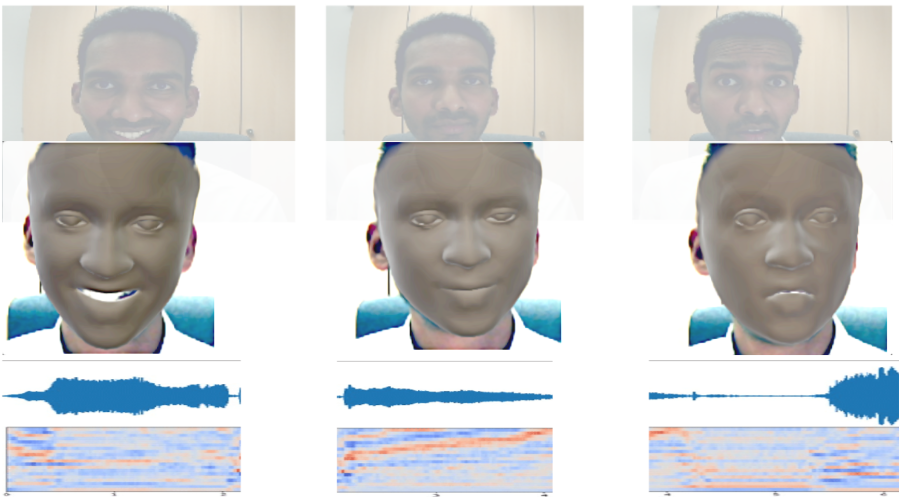}
    \caption*{Speaker: Multi-modal social cues}
    \label{fig:sub1}
  \end{subfigure}
  \begin{subfigure}[b]{0.35\linewidth}
    \includegraphics[width=\linewidth]{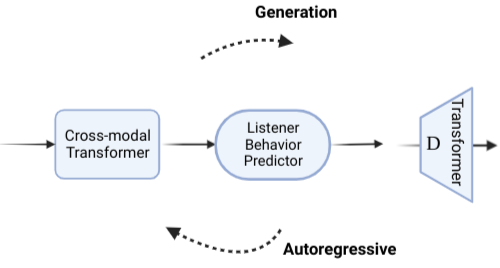}
    \caption*{listener behaviour prediction}
    \label{fig:sub2}
  \end{subfigure}
  \begin{subfigure}[b]{0.25\linewidth}
    \includegraphics[width=\linewidth]{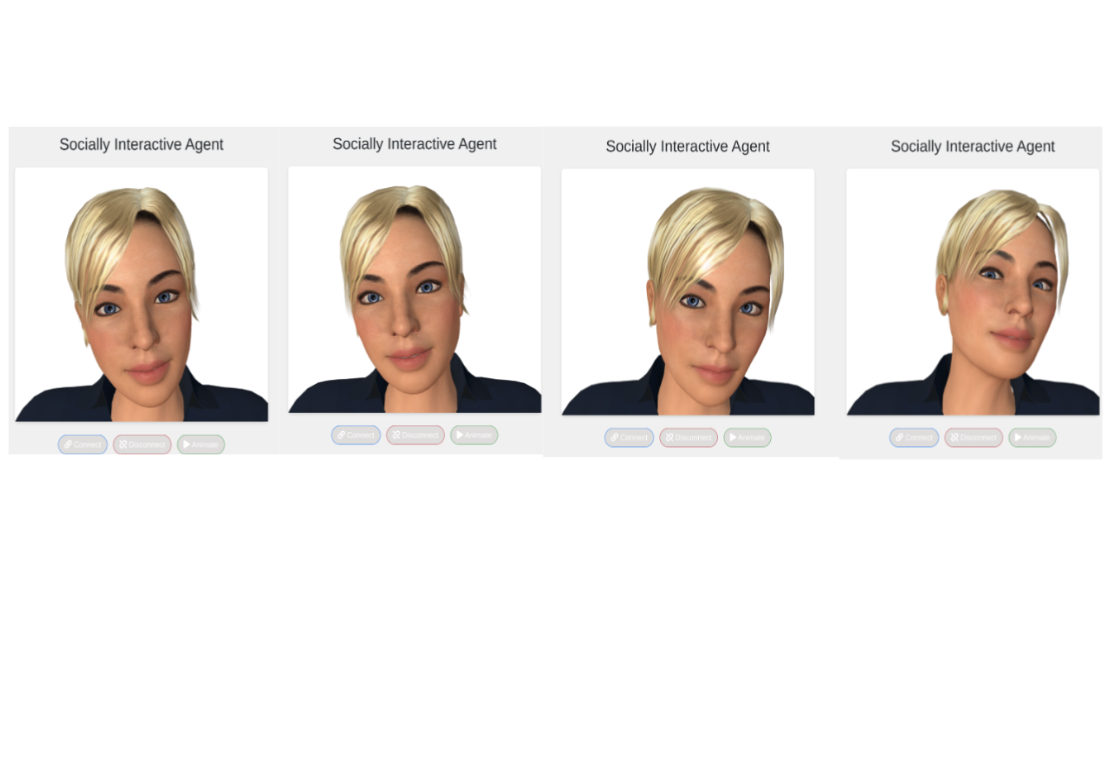}
    \caption*{Listener: Interactive virtual agent}
    \label{fig:sub3}
  \end{subfigure}
  \caption{Experience ReNeLiB: A framework transforming human-webcam interactions through 3D motion and Mel frequency analysis, enabling interactive virtual agents to adapt to user behavior based on multimodal social cues.}
  \label{fig:teaser}
\end{teaserfigure}

\maketitle

\section{Introduction}
\label{section:Introduction}
Socially-aware Interactive Agents (SIAs) are autonomous systems proficient in engaging in natural language dialogues and interacting with their environment~\cite{multimodalBehaviousforSIA}. The increasing importance of human-machine interaction in everyday life necessitates the development of SIAs that can actively listen and respond to users in a believable and context-dependent manner. The Media Equation theory posits that individuals treat computers, televisions, and new media similarly to real people and places~\cite{themediaequation}, suggesting that SIAs displaying social behaviors, including active listening, can positively impact user experiences. It is essential to argue that SIAs should operate in a manner tailored to the context in which they are deployed. This emphasizes the need for SIAs to demonstrate context-sensitive, believable, and professional behavior. By doing so, these agents can create more immersive and natural interactions, ultimately leading to improved human-SIA communication and user satisfaction.

Facial expressions and head movements are essential components of human communication and social interaction, extending their importance to human-SIA interactions. Bickmore and Cassell ~\cite{bickmore1999small} evaluated an embodied conversational agent (ECA) that used nonverbal cues, such as nodding and eyebrow movements, to express active listening. Their findings revealed that users perceived the ECA as more attentive and engaging compared to a version without these social cues. Context comprehension and accounting for the multimodal nature of interactions are crucial, as facial gestures and spoken utterances are intrinsically interconnected. For instance, in psychotherapy, nonverbal synchrony has proven vital, with head synchrony positively correlating to therapy success~\cite{ramseyer2014nonverbal}. Consequently, incorporating context-dependent facial gestures in active communicative listening behavior is imperative for fostering natural human-SIA interactions.

Traditional authoring and assistive frameworks for SIAs have often relied on predefined scripts to generate non-verbal behavior, playing a crucial role in the development of conversational agents and virtual characters. For example, the BEAT toolkit enables the automatic generation of gestures and facial expressions based on input text~\cite{beats}. Moreover, the Behavior Markup Language (BML) provides a unified framework for generating and controlling multimodal behavior in virtual agents, including facial expressions, gestures, and gaze direction~\cite{kopp2006towards}. Additionally, Pelachaud's work on modeling multimodal emotional expression in virtual agents has contributed to the understanding of how predefined scripts can be used for generating non-verbal behaviors~\cite{pelachaud2009modelling}. However, these methods may be limited in capturing the natural complexity and dynamic nature of human-human non-verbal communicative behavior, necessitating further research on alternative techniques that can address these limitations.

The advent of deep learning techniques has facilitated data-driven generative approaches for creating locomotion, dancing, and facial gestures in avatars~\cite{Jonell2020,ng2022learning2listen}. Despite their potential, these techniques exhibit limited real-time dyadic capabilities, rendering them unsuitable for direct implementation in SIAs. The importance of real-time capabilities in practical applications is critical for fostering immersive and engaging human-SIA interactions~\cite{thies2020face2face}. As such, there is a pressing need to address these limitations and develop a framework using data-driven generative methods that can meet the real-time dyadic requirements of SIAs, enhancing the naturalness and efficacy of human-SIA communication across various contexts.


In this paper, we present a novel open-source modular software toolkit designed to overcome these limitations by facilitating the integration of data-driven behavior generation with multiple facial parametric representations. Our approach employs the FLAME (Faces Learned with an Articulated Model and Expressions)~\cite{FLAME:SiggraphAsia2017}, a highly expressive 3D face parametric representation. To accommodate other facial parametric representations, we devise a mapping function for Apple Inc.'s ARKit ARFaceAnchor~\cite{appleartoolkit}. Our toolkit incorporates state-of-the-art generative models for real-time non-verbal active listening behaviors in dyadic SIA interactions, dynamically adjusting listener behavior according to conversational context. This framework bridges the gap in existing approaches by enabling real-time, multimodal feature representation and seamless integration of data-driven models. Furthermore, we present a method to enhance the expressiveness of industrial-standard SIAs using the  commercial VuppetMaster platform, developed by Charamel~\cite{VuppetMa82:online}. This is achieved by establishing a coherent transformation between FLAME expression coefficients and Apple ARKit expressions, enabling seamless integration of generated behavioral expressions within the VuppetMaster platform. By offering an open-source framework for data-driven listener behavior generation, our work paves the way for the development of increasingly sophisticated SIAs and their applications in a wide range of contexts. 



\section{Related Work}
\label{section:RelatedWork}
Our work is related to interactive virtual agents, recent approaches to data-driven behavior generation, and representation systems for human visual social behavior.

\subsection{Interactive Virtual Agents}
Interactive Virtual Agents (IVAs) aim to generate dynamic social behaviors and maintain user engagement in real-time, fully dyadic conversations~\cite{SARA, babyxco, Gebhard2012}. The significance of interactional motion within conversational agents has been increasingly recognized, as it enhances user engagement and fosters more natural communication experiences. Studies have investigated rapport building in virtual agents~\cite{virtualRapport, virtualraport2}, the impact of animated conversations on user experience~\cite{animConversation, SARA}, and the importance of situated interaction in IVAs~\cite{BohusDan}. Additionally, Gebhard and colleagues~\cite{Gebhard2012} explore the role of gestures and body language in improving communication, proposing a conversational flow for real-time, fully dyadic interactions.

Nonetheless, the majority of prior research has focused on rule-based methods, employing motion capture sequences or hand-crafted animations for interactional motion in facial gestures and speech~\cite{Gebhard2012, BohusDan, SARA}. These rules encompass gaze behavior~\cite{gazerule}, turn-taking management~\cite{virtualraport2}, facial expressions~\cite{FacialExpressionsRule}, gestures and body language~\cite{beats}, and backchanneling~\cite{backchannelingRule}. These approaches exhibit limitations regarding the range of captured gestures and the simplifying assumptions made for motion generation, rendering them less suitable for context-dependent dynamic interactions.

\subsection{Data-driven Approaches for Behavioral Animation Synthesis}

In response to the limitations of rule-based methods, recent studies have explored data-driven approaches for generating conversational behavior in IVAs while leveraging large datasets and advanced modeling techniques to capture the subtleties of human behavior. For example, \cite{Kucherenko2020} investigated machine learning techniques and deep learning models to create more contextually relevant and natural speaker behaviors. Extending the automatic locomotion synthesis framework MoGlow~\cite{Henter2020}, the "Let's face it"~\cite{Jonell2020} study devised a probabilistic approach for synthesizing facial gestures that account for interlocutor awareness in dyadic conversations. However, this method's efficacy was constrained by not differentiating between speech-related and non-speech-related facial gestures during feature extraction. In light of the interdependence between speech and gesture perception, the "Learn2Listen" method employed transformer-based VQ-VAE and multimodal fusion techniques~\cite{ng2022learning2listen} to predict non-verbal facial behavior, yielding promising results in capturing facial gesture nuances and efficient behavior generation. Despite these advancements, the integration of data-driven approaches into real-time, fully dyadic conversational flows remains a challenge due to the lack of an appropriate framework capable of handling the computational requirements, synchronizing multi-modal inputs, and facilitating seamless integration with existing IVA systems~\cite{Gebhard2012, babyxco}.

\subsection{Behaviour Representation}

Advancements in photorealistic virtual humans have led to more believable and engaging representations, significantly contributing to the development of intelligent virtual agents~\cite{souleMachine:online}. A primary challenge in this domain, however, is the lack of standardization in agent animation, which impedes progress in automatically generating realistic agent behavioral animation using data-driven approaches. Different systems utilize various techniques, such as blendshape and bone animation, to define their animation controllers~\cite{kopp2006towards}. Additionally, creating photorealistic virtual humans for real-time IVAs demands considerable expertise and resources due to the process's inherent complexity~\cite{advanceHumanModel}.

Van der Struijk and colleagues~\cite{facsavatar} employed the 3D human model from the open-source FACSHuman\footnote{\url{https://www.michaelgilbert.fr/facshuman/}} software add-on to drive facial motors in real-time using the Facial Action Coding System (FACS) detected with OpenFace~\cite{openface}. However, their approach focused on mimicking facial behavior rather than generating social behavior based on the interactive context in Human-Computer Interaction setups using IVAs. A limitation of using FACS for facial action unit representation is its inability to effectively capture subtle expressions and head rotations~\cite{Jonell2020}. Furthermore, \cite{facsavatar} highlighted the limitations of OpenFace~\cite{openface} in detecting FACS Action Units (AUs) and intensity values, as its AU and intensity value predictors are not synchronously trained, leading to inaccuracies. As a result, manual post-processing was necessary in~\cite{facsavatar} to fine-tune the intensity value for activating facial animation.

ARFaceAnchor\footnote{\url{https://developer.apple.com/documentation/arkit/arfaceanchor/blendshapelocation}} has been developed to enable real-time face tracking systems on native devices. Projects such as~\cite{MetaHuma26:online} have utilized ARKit to animate socially interactive agents in real-time. However, using ARKit introduces significant drawbacks due to its device dependencies, as the facial 3D mesh cannot be employed outside the native platform. This limitation hinders the extraction of facial expressions and head movements from large video datasets recorded with monocular cameras, which are essential for training generative machine learning models like~\cite{Jonell2020, ng2022learning2listen}.

Researchers have explored open-source alternatives such as RingNet~\cite{RingNet:CVPR:2019}, which learns to regress 3D face shape and expression from an image without 3D supervision, offering functionality comparable to ARFaceAnchor yet with different animation controllers. However, RingNet's complex neural network architecture results in substantial computational expense~\cite{ng2022learning2listen}. DECA~\cite{DECA}, an improved version of RingNet, leverages the FLAME model~\cite{FLAME:SiggraphAsia2017} and a convolutional neural network for efficiently capturing and animating 3D facial expressions from single 2D images. Despite its ability to generate more realistic facial reconstruction, real-time processing remains challenging due to DECA's face-alignment module, which causes a computational bottleneck. EMOCA~\cite{EMOCA:CVPR:2021} extends DECA's implementation to improve 3D facial reconstruction with higher emotional fidelity, employing a deep perceptual emotion consistency loss during training. This novel approach outperforms existing methods in expression quality and perceived emotional content, demonstrating the potential of 3D geometry, yet it has not been extended in reconstructing 3D facial representations for real-time face tracking and 3D morphing. We extended this method to work in real-time 3D morphing and develop a novel facial expression transformation, that aims to bridge the gap between data-driven techniques and commercially available IVAs, fostering seamless integration and improved human-computer interaction experiences.

\newcommand{\shapecoeff}{\beta}
\newcommand{\posecoeff}{\theta}
\newcommand{\expcoeff}{\psi_{e}}

\section{Framework}
\label{section:concept}

Human-to-human communication involves a complex interplay of verbal and nonverbal cues. Furthermore, the behavior of a listener during a conversation can depend on the context of the conversation and the social setting, as evidenced by the nuances observed in human-to-human communication and real-life therapeutic interactions, as shown in Fig.~\ref{fig:teaser}. 

Our main objective is to develop a framework that predicts a socially interactive agent's listening facial behavior in real-time based on the user's multimodal social cues. Specifically, we aim to predict the interactive agent's facial expressions, $\mathbf{\hat{F}}^{sia}_t$, at each time-step $t$, given the user's contextual speaking information encapsulated in audio features $\mathbf{A}^{user}_{1:t}$ and facial features $\mathbf{F}^{user}_{1:t}$, and any past predicted facial behavior for the agent to process them in an autoregressive manner, predict $w$ listener behavioral sequence $\mathbf{{F}}^{sia}_{t:t+w}$. Therefore, we model the distribution $P$ of the agent's predicted facial behavior, learned from the therapist's listening behavior $\mathbf{\hat{F}}^{thera}_t$ conditioned on audio features from the patient $\mathbf{A}^{pat}_{1:t}$ and facial features $\mathbf{F}^{pat}_{1:t}$ taking into account the patient's multimodal contextual information. Therefore, we model the distribution $P$ of the interactive agent's predicted listening behavior, learned  $\mathcal{L}$ from the therapist's listening behavior, as:

\begin{equation}\label{eq:objective}
\begin{aligned}
P(\mathbf{\hat{F}}^{sia}_t \mid \mathbf{A}^{user}_{1:t}, \mathbf{F}^{user}_{1:t}) = \mathcal{L}(P(\mathbf{\hat{F}}^{thera}_t \mid \mathbf{A}^{pat}_{1:t}, \mathbf{F}^{pat}_{1:t})) 
\end{aligned}
\end{equation}

Our framework then utilizes $P$ distribution for predicting a socially interactive agent's (SIA) listening facial behavior with the user's multimodal social cues processed in real-time. The framework employs deep learning techniques to model the relationship between user's audio and facial features and the agent's facial behavior, using a dataset of real-life therapeutic video data recordings for training. Comprising several interconnected modules, such as the User with Voice Activity Detector, Real-time Feature Extraction, Behavior Generator, FASTApi Server with Local-to-Global Transformation, and LiveFLAME Blender Add-on, the framework is designed for high throughput and real-time processing supporting both FLAME-based and ARKit-based interactive agents. By leveraging the ZeroMQ\footnote{\url{https://zeromq.org/}} distributed messaging library and web sockets for communication, the framework aims to enhance the expressiveness and responsiveness of listening behavior for the SIA in various applications, such as telemedicine, mental health counseling, and customer support.

\subsection{Framework Modules}\label{sec:frameworkmodules}


\begin{figure*}[ht!]
    \begin{center}
        \includegraphics[width=\linewidth]{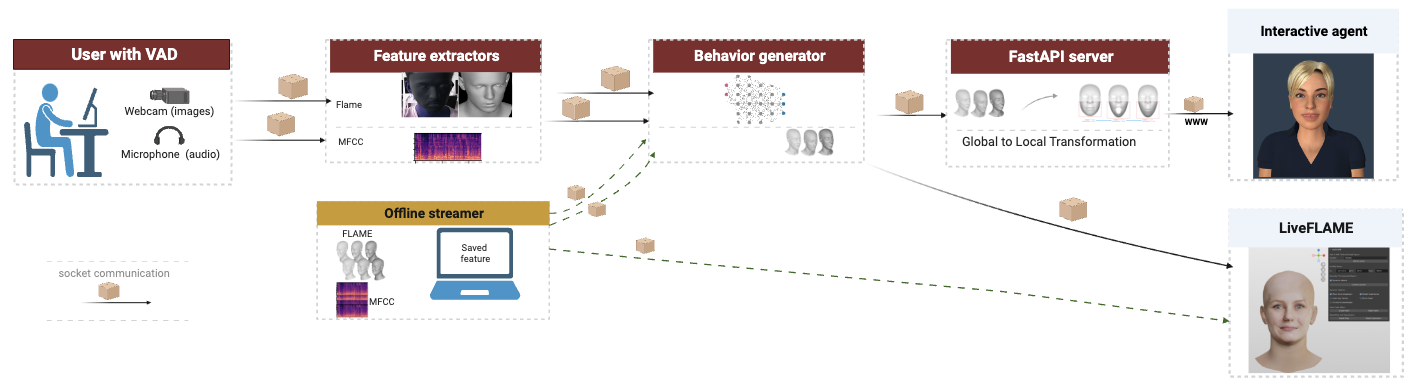}
    \end{center}
 \caption{\textit{Overview of the real-time interactive framework modules, utilizing a publisher-subscriber pattern to enable real-time listener behavior. The framework supports both online and offline modes, allowing for feature extraction from sensors or streaming from locally saved files. The solid arrows represent the modules used in our evaluation.}}
  \label{fig:frameworkOverview}
\end{figure*}

The proposed framework, depicted in Fig.~\ref{fig:frameworkOverview}, comprises several interconnected modules that collaborate to generate the listening facial behavior of a SIA in real-time. The framework has been designed to support both online processing, using sensor data, and offline processing, which allows from prior extracted feature data. In this section, we present a comprehensive description of each module. The modules work together to enable the SIA to exhibit realistic facial behavior. 

\begin{enumerate}[leftmargin=*,labelindent=0pt]
\item \textbf{User with Voice Activity Detector} is responsible for detecting when the user is speaking. Recognizing voice activity activates the listener behavior for the SIA, enabling more natural and responsive interactions between the user and the agent.

\item \textbf{Real-time Feature Extraction} captures and processes audio and facial features, specifically FLAME and MFCC features, synchronously. Proper alignment between the extracted features is ensured using timestamps, which helps maintain accurate temporal information for the input data.

\item \textbf{Behavior Generator} employs a producer-consumer pattern using fixed-length double-ended queues for efficient data handling in a multi-threaded environment. This design allows multimodal producers to write data to the queues, while a consumer thread processes the data within a sliding window, predicting facial behaviors with an adjustable processing rate and publishing the data to the subsequent module in the pipeline.

\item \textbf{FASTApi Server with Local-to-Global Transformation} serves as the back-end for the web-based IVA front-end. It streams data to a IVA such as VuppetMaster\footnote{https://www.charamel.com/en/software/vuppetmaster} or MetaHuman \cite{MetaHuma26:online}, processing rotational and facial expression transformations such as FLAME to ARKit.

\item \textbf{Interactive Virtual Agent} is a web-based plugin powered by VuppetMaster, designed for the integration into HTML web browsers. This could be adjustable to work with another character animator like MetaHuman~\cite{MetaHuma26:online}.

\item \textbf{LiveFLAME Software Add-on} is a real-time visualizer of FLAME parameters within the Blender\footnote{\url{www.blender.org}} software. Implemented as a Blender add-on, it enables users to monitor and evaluate the generated facial behavior by directly mapping the predicted FLAME parameters to a FLAME-based facial model (e.g., female or male). This visualization tool provides valuable insights during the development and testing of the framework.

\end{enumerate}


\section{Implementation}
\label{section:realization}

In this section, we discuss the implementation of the proposed system, also including an evaluation of the trained behavior generative models and the dataset utilized.

\subsection{Facial and Audio Representations}
\label{sec:facialandaudiorep}

Our framework employs the FLAME statistical 3D head model~\cite{FLAME:SiggraphAsia2017} and the EMOCA face reconstruction framework \cite{EMOCA:CVPR:2021} to represent facial expressions and head movements. The FLAME model consists of three critical components: expression, pose, and shape vectors. The expression vector $\expcoeff \in \mathbb{R}^{1\times 100}$ captures variations in facial expressions, encoding facial muscle movements within a reduced-dimensional space derived from a 3D scanned dataset using Principal Component Analysis (PCA). The pose vector $\posecoeff \in \mathbb{R}^{1\times 15}$ represents head orientation and rotations of specific joints (e.g., neck, jaw, eyeballs), describing the overall position and orientation of the 3D head model in 3D space. The shape vector $\shapecoeff \in \mathbb{R}^{1\times 300} $ encodes individual facial identity, encompassing the unique structure and geometry of the face. This vector defines the base 3D head model, which is subsequently modified by the expression and pose vectors to create the final 3D head representation.


For prosodic behavior modeling, we extracted Mel Frequency Cepstral Coefficients vector $\mathcal{MFCC} = { c_1, c_2, ..., c_l }$ from the audio as an audio representation, where l represents the number of coefficients. MFCCs capture phonetic information and provide a compact representation of the audio signal, making them suitable for real-time applications with compact data footprints for social cues via audio. Moreover, MFCCs have been extensively employed in speech and emotion recognition tasks \cite{mfccSchuller}, demonstrating their efficacy in capturing relevant information from audio signals.

\subsection{\textbf{Real-life Therapy Interaction Dataset}} \label{sec:therapyinteraction}
\begin{figure*}[h!]
\begin{center}
\includegraphics[width=\linewidth]{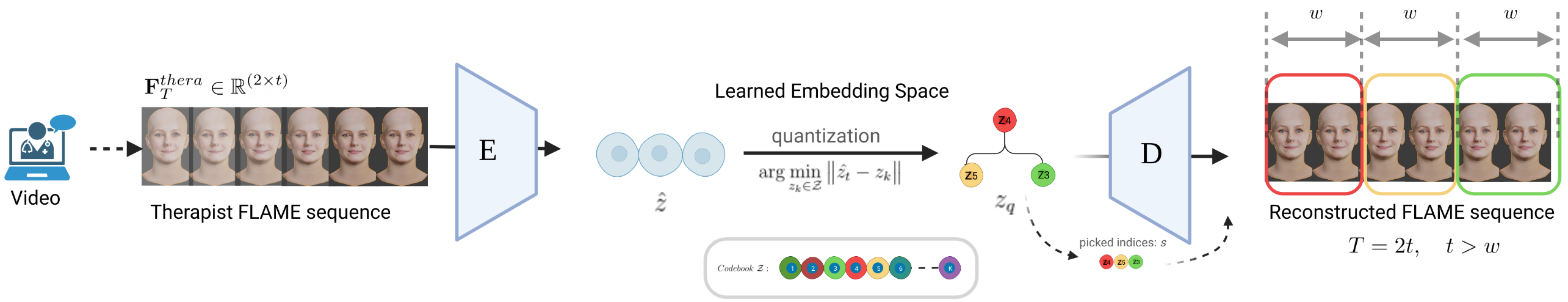}
\end{center}
\caption{\textit{VQ-VAE training process}}
\label{fig:vqvae}
\end{figure*}
In recent years, the development of virtual therapists has garnered significant interest, with the aim of providing mental health support through digital platforms. However, obtaining access to real-life therapeutic video data, to train data-driven generative models, remains a challenge due to the sensitive nature of such interactions. To address this issue, we collaborated with \cite{Peham2015, opd} to acquire real-life therapeutic video data recordings. Further details about the data are listed in Appendix \ref{appendixA}.

Given the sensitive nature of patient-therapy interactions, it is crucial to ensure the confidentiality of the data. As such, we employed a Secure Machine Learning Architecture (SEMLA)~\cite{semla:online} for data processing and machine learning model training. Our primary objective was to process separate video streams and audio channels for both patients and therapists, in order to create a feature dataset for model training. 


\subsection{Conditional Motion Synthesis of Conversational Dynamics}

This section outlines the training and evaluation processes of unsupervised machine learning models designed for conditional motion synthesis between a speaker and a listener. The approach by Ng et al.~\cite{ng2022learning2listen} serves as our foundation. We extend the original method in two significant ways. Firstly, we replace the DECA 3D Morphable Face Model~\cite{DECA} with EMOCA~\cite{EMOCA:CVPR:2021} to estimate 3D facial expressions. Secondly, we used the pyanote speaker-diarization method\footnote{\url{https://github.com/pyannote/pyannote-audio}} for identifying and separating speakers during interactions. Following the methodology in~\cite{ng2022learning2listen}, our learning task is represented in Eq.~\ref{eq:objective}. We then proceed with model training. 


\subsubsection{\textbf{Model Training}}\label{sec:model_training}
Data were derived from sessions conducted by "TherapistA". This segmentation resulted in intervals: \(S_{\text{backchanneling}}\), \(S_{\text{short-speech}}\), and \(S_{\text{long-speech}}\). Additional details regarding speech activity segmentation are presented in Appendix \ref{appendixB}. The TherapistA dataset spans 12 hours, recorded at 25 fps, presenting split video recordings with the therapist positioned to the right and the patient to the left. The model training process encompassed an initial phase of VQ-VAE pre-training and the Predictor module is trained seperately. The Predictor model is excluded from backpropagation during E, Z, and D training. The notation employed in this section is consistent with that found in~\cite{ng2022learning2listen}.

\begin{figure}[ht!]
\begin{center}
\includegraphics[width=\linewidth]{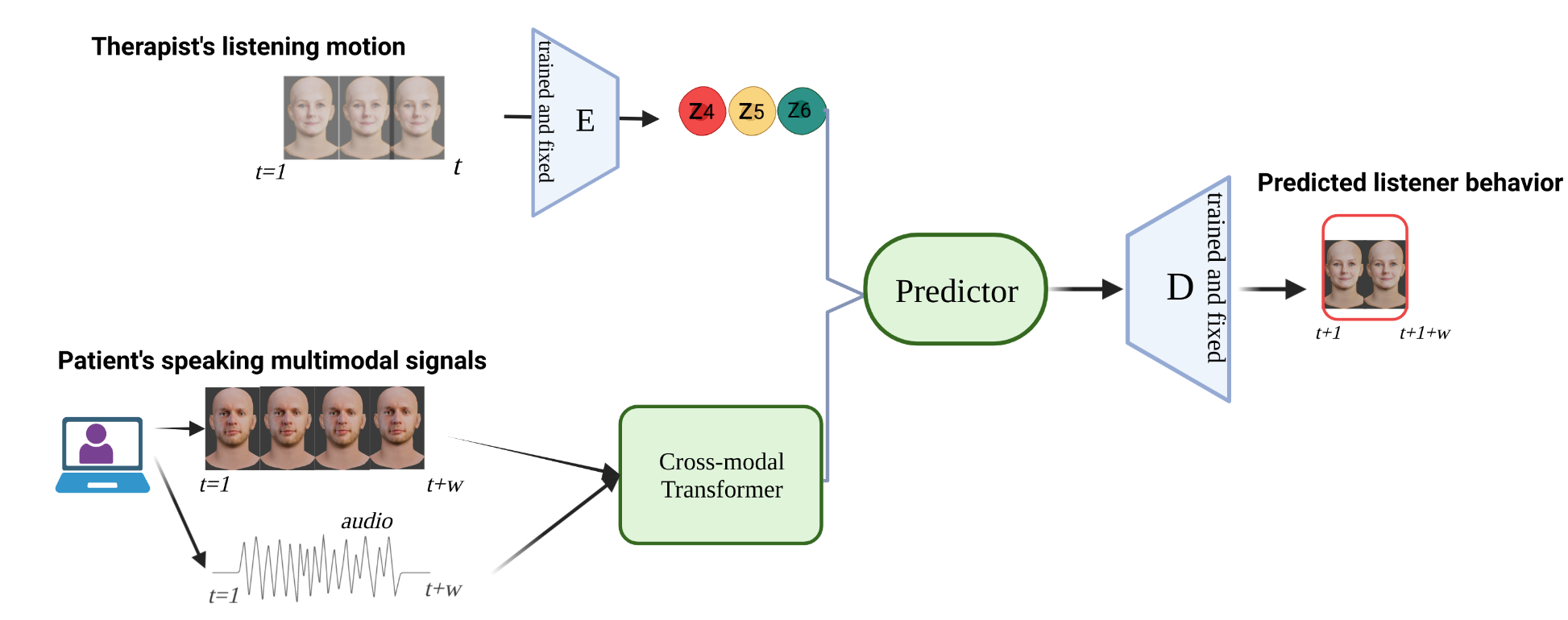}
\end{center}
\caption{\textit{Predictor training process}}
\label{fig:predictor}
\end{figure}

\textit{\textbf{Pre-training the VQ-VAE}}: Our approach employs a transformer-based encoder (E) and decoder (D) architecture to efficiently capture the therapist's facial motion and expression from video data, as illustrated in Fig.~\ref{fig:vqvae}. The training process involves the encoder, decoder, and codebook (Z) components, utilizing the loss function described in \cite{ng2022learning2listen}. The dataset is partitioned into 70\% training, 20\% validation, and 10\% testing portions. The Adam optimization algorithm is used to fine-tune the model's parameters with the goal of minimizing the loss function. After approximately 5,000 training steps, the best-performing model on the validation set is preserved for the next phase.

\textit{\textbf{Training the Predictor Module}}: Upon pre-training the VQ-VAE, the encoder and codebook components are kept fixed, and the focus shifts to training the transformer-based predictor module, as depicted in Fig.~\ref{fig:predictor}. The predictor is designed to learn temporally long-range patterns in the input sequence by employing cross-modal attention to fuse audio features and facial motion features as the conditioning vector. This process is coupled with the discretized past listener motion sequence encoding provided by the pre-trained encoder. The autoregressive predictor outputs a distribution over the $K=200$ discrete codebook indices, from which a code for the subsequent timestep is sampled and then passed to the trained decoder. We selected values $T=64$, $t=32$, and $w=8$, along with a Mel frequency length of $l=128$, similar to \cite{ng2022learning2listen}, to evaluate our model with their pre-trained model.

We applied the model trained on TV interviewer Conan, as established by \cite{ng2022learning2listen}, which encompassed a range of participants in interviews. To evaluate the TherapistA model, a comparative matrix was devised, incorporating both its ground truth data and the ground truth from interviewees interacting with Conan. The same assessment was performed for the Conan model using the ground truth data of patients and interviewees. We calculated L2 loss to ensure compatibility with the reference study \cite{ng2022learning2listen} while their baseline model achieved an L2 loss of 52.68.

\begin{table}[ht!]
    \centering
    \caption{L2 Loss values for machine learning models tested on different datasets}
    \footnotesize 
    \setlength{\tabcolsep}{3pt} 
    \begin{tabular}{lcc}
        \toprule
        & \multicolumn{1}{c}{Therapist} & \multicolumn{1}{c}{Interviewer} \\
        \cmidrule(lr){2-2} \cmidrule(lr){3-3}
        Dataset & L2 Loss & L2 Loss \\
        \midrule
        Patients & \textbf{41.06} & 89.49 \\
        Interviewee   & 78.85 & \textbf{59.72} \\
        \bottomrule
    \end{tabular}
    \label{tab:loss_values}
\end{table}

\subsubsection{Discussion}
It is important to recognize that the L2 loss accentuates disparities between predicted and ground truth temporal FLAME vector sequences, leading to larger error values for substantial deviations. The results, presented in Table \ref{tab:loss_values}, should be interpreted with the understanding that there are language and interaction context differences between the two datasets. This highlights the significance of behavior modeling within the interaction context. 


\subsection{Real-time Framework Application}


\begin{figure}[ht!]
\begin{center}
\includegraphics[width=\linewidth, height=5cm]{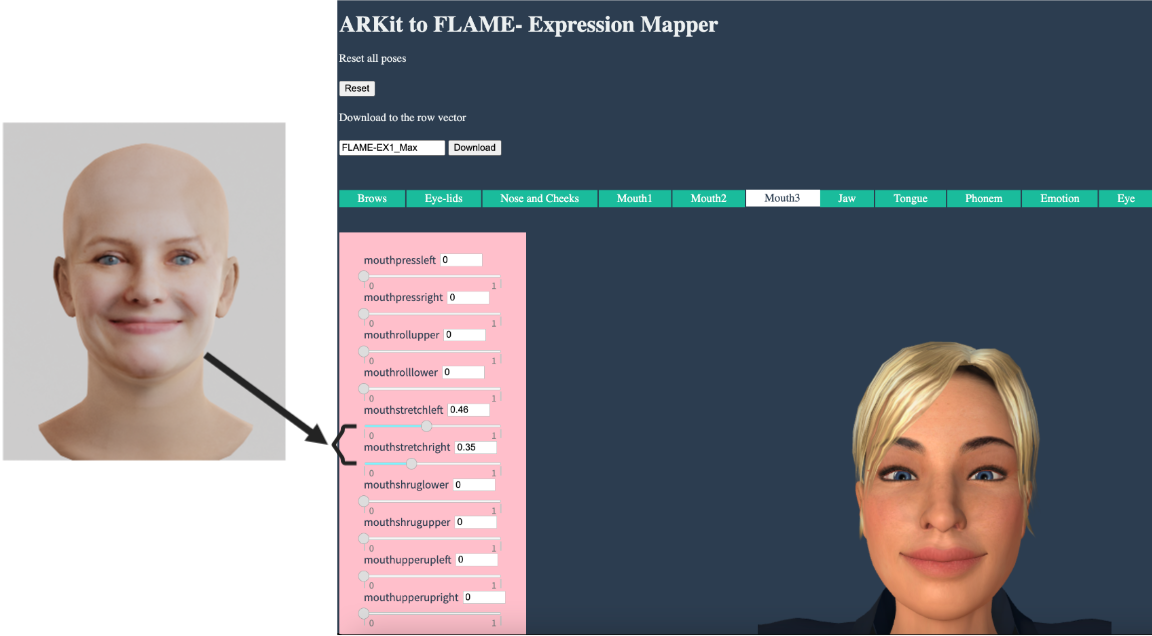}
\end{center}
\caption{\textit{FLAME to ARKit expression mapper}}
\label{fig:expressionMapper}
\end{figure}

The models that we developed predict behavior in the FLAME head-pose and expression format, necessitating a FLAME-compatible interactive agent for visualization. We utilize the LiveFLAME software addon as described in Section \ref{sec:frameworkmodules}. However, given the rapid advancements in photorealistic socially interactive agents \cite{MetaHuma26:online, souleMachine:online, VuppetMa82:online}, each with their unique facial animation systems, our objective is to enhance the compatibility of our synthesized behaviors with industry-standard socially interactive agents, such as VuppetMaster~\cite{VuppetMa82:online}. Notably, Charamel \cite{VuppetMa82:online} supports the research community by providing IVAs for collaborative projects. VuppetMaster's visually appealing full-body agents (see the agent in the upper right of Figure 3) are animated using VuppetMaster's animation engine and deployed as a web-based solution. This engine is based on a humanoid facial skeletal structure incorporating facial muscle controllers, which enable a comprehensive range of facial expressions and movements. The controllers' naming scheme and shape manipulation capabilities are designed to be compatible with ARKit facial muscle controllers \cite{appleartoolkit}. However, it should be noted that the facial muscle controllers in the FLAME mesh follow a distinct representation, necessitating a suitable conversion process for seamless integration.

\subsubsection{Constructing the Global-to-Local Transformation Matrix}
\label{sec:global_to_local}

In this work, we introduce a novel real-time linear transformation designed for efficient execution in real-time applications. While the FLAME model employs global expressions, simultaneously controlling multiple facial muscles, ARKit utilizes local expressions that focus on individual muscle control, enabling more localized facial movements. We aim to develop a computationally efficient transformation matrix for real-time scenarios. To facilitate the transformation between FLAME expression coefficients and Apple ARKit facial expressions, a global-to-local transformation matrix $\mathcal{GL}$ is constructed. Furthermore, the jaw and head poses are independently converted from axis angle representation to Euler angles.

The row vectors of $\mathcal{GL}$ are derived by mapping extreme FLAME expressions using multiple  corresponding ARKit expressions using an expression mapper, as shown in Fig.~\ref{fig:expressionMapper}. This process is iterated for selected $\pm{F_e} \in \mathbb{R}^{ \left\lVert \expcoeff \right\rVert \times 2}$ FLAME expressions, where the "+-" in the equation indicates the extremes (-3 and 3) of each FLAME expression.


\begin{equation}
\mathcal{GL} = \begin{bmatrix} \vdots & \vdots & \vdots & \ddots & \vdots \\
\mathbb{\pi}_e(i,1) & \mathbb{\pi}_e(i,2) & \mathbb{\pi}_e(i,3) & \cdots & \mathbb{\pi}_e(i,52)
\end{bmatrix}
\end{equation}

\subsubsection{Real-Time Transformation of FLAME Vectors to ARKit Vectors}
\label{sec:flame_to_arkit}

Given a batch size $n$ of FLAME vectors $\mathbf{\mathcal{F}}$, the transformed to $n \times 52$ ARKit expression matrix  $\mathbf{\mathcal{A}}$ computed using the tranformation matrix:

\begin{equation}\label{eq:tranformation_clip}
\mathcal{A}{(n \times 52)} = \text{normalise}(\mathcal{F}{(n \times F_e)} \times \mathcal{GL}{(F_e \times 52)}, 0, 1)
\end{equation}

For the conversion of rotation angles, the following operations are performed:

\begin{flalign}
\begin{split}
\boldsymbol{q}_{\text{jaw}} = \text{axisAngleToQuaternion}(\boldsymbol{\alpha}_{\text{jaw}}), \\
\boldsymbol{q}_{\text{head}} = \text{axisAngleToQuaternion}(\boldsymbol{\alpha}_{\text{head}}), \\
\boldsymbol{e}_{\text{jaw}} = \text{quaternionToxyzEuler}(\boldsymbol{q}_{\text{jaw}}), \\
\boldsymbol{e}_{\text{head}} = \text{quaternionToxyzEuler}(\boldsymbol{q}_{\text{head}}),
\end{split}
\end{flalign}

\section{EVALUATION}
\label{section:evaluations}


\begin{figure*}[ht!]
  \centering
  \begin{subfigure}[b]{0.4\textwidth}
    \includegraphics[width=\textwidth,height=4cm]{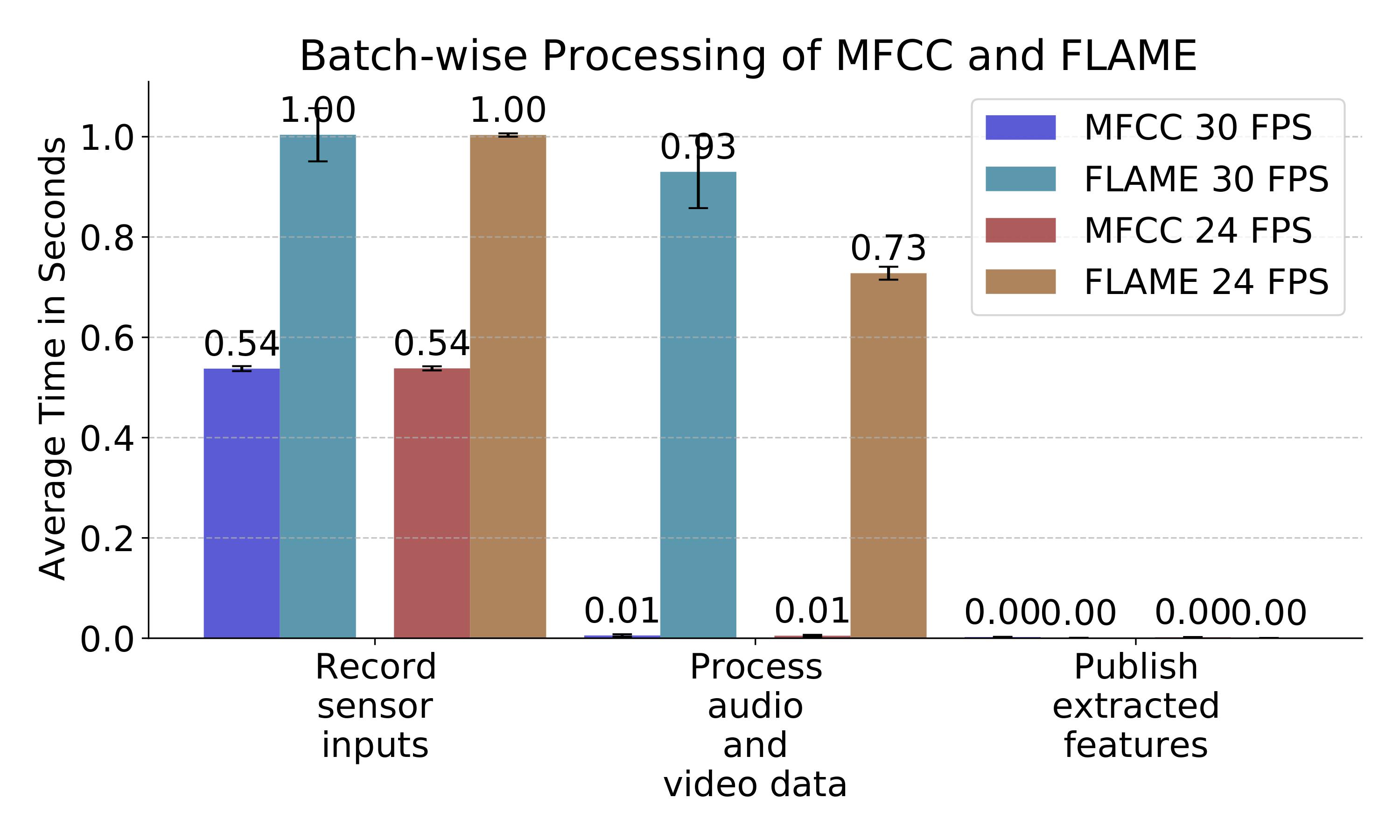}
    \caption{Feature extractors in parallel}
  \end{subfigure}
  \begin{subfigure}[b]{0.4\textwidth}
    \includegraphics[width=\textwidth, height=4cm]{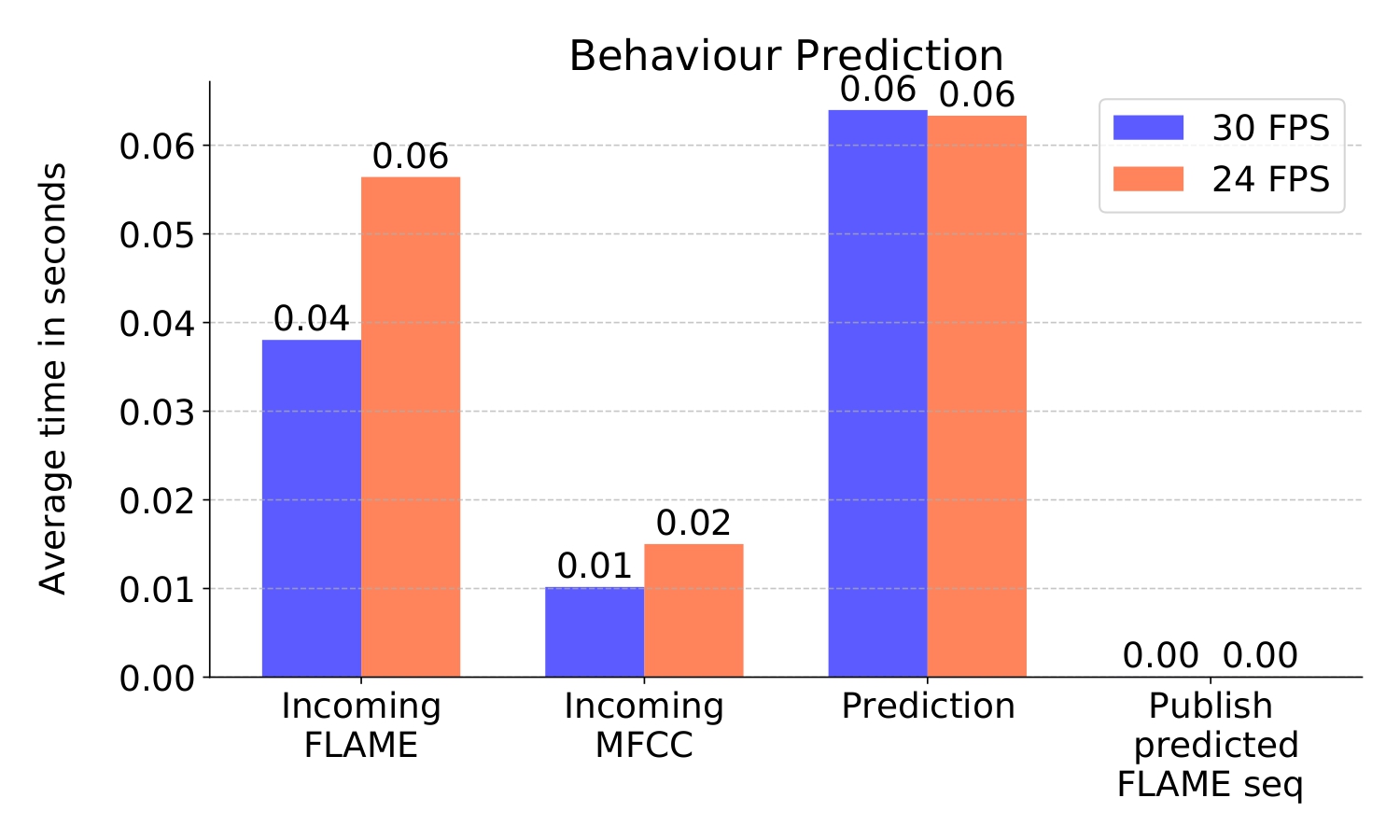}
    \caption{Behaviour predictor}
  \end{subfigure}
   \begin{subfigure}[b]{0.4\textwidth}
    \includegraphics[width=\textwidth,height=4cm]{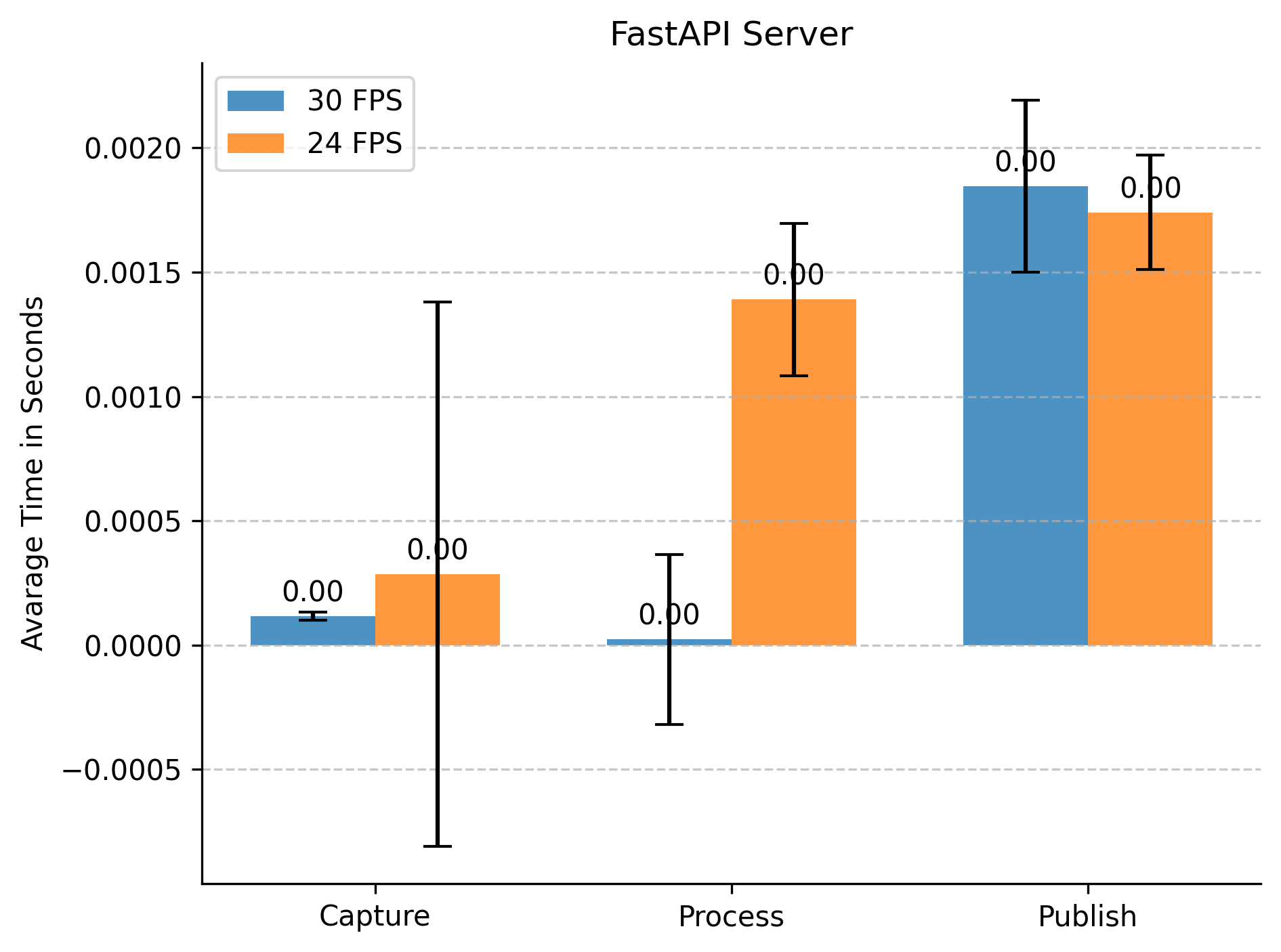}
    \caption{FLAME to ARKit GL transformation}
  \end{subfigure}
  \begin{subfigure}[b]{0.4\textwidth}
    \includegraphics[width=\textwidth,height=4cm]{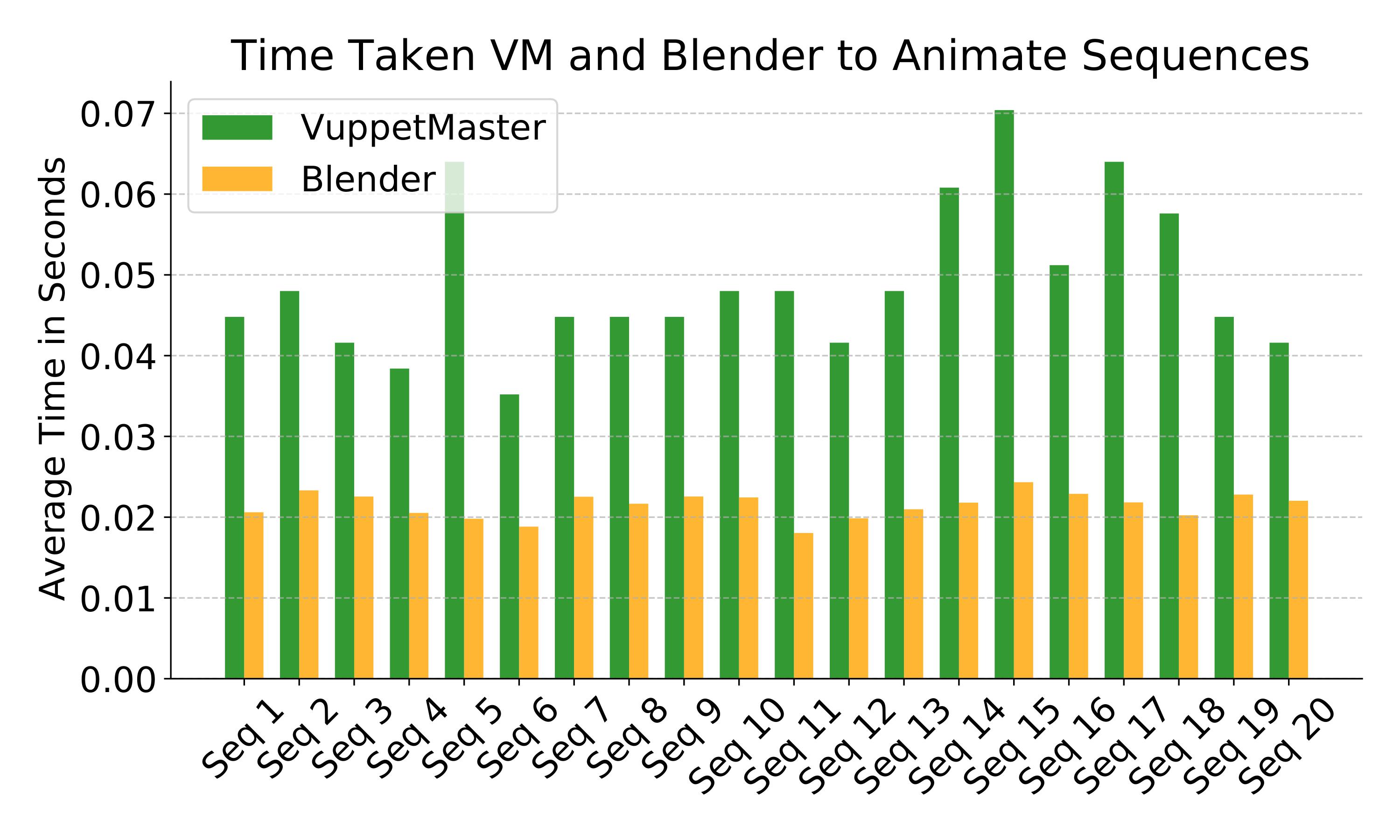}
    \caption{IVA animation  at 30 fps}
  \end{subfigure}
  \caption{Network latency plot for each framework module for different frame rates}
  \label{fig:realtime}
\end{figure*}

In our system, real-time operation is achieved through parallel processing and a modular architectural approach. We adopt the runtime evaluation metrics proposed by "Facsvatar" \cite{facsavatar}. Video features, represented by FLAME, and audio features, denoted by MFCCs, are extracted concurrently. This simultaneous extraction ensures seamless streaming of features using ZeroMQ, with timestamps to guarantee synchronization. We conducted evaluations of our system on both high-end (13th generation Intel Core i9 and GeForce RTX 4090) and mid-range (Intel Core i7 and RTX 1080) configurations. Notably, while individual modules, such as Mel and FLAME for behavior generation, demand processing time, the system's intrinsic parallelism significantly mitigates the latency from a user's input to the agent's response. In this study, we present results from a test scenario conducted on an i9 platform running Ubuntu 22.04. We thoroughly assessed the performance of the ReNeLiB framework across various components.



In our design, each module employs a multi-threaded custom producer-consumer network architecture, using either ZeroMQ or WebSocket, to ensure independent data processing. This architecture utilizes a publisher-subscriber communication pattern, optimizing module-specific concerns and enabling flexible data processing speeds. The MFCC extractor adjusts to audio recording duration $T_{audio}$ and sampling rate $R$, and publish the data to the behaviour predictor at $M_{fps}$. Concurrently, the FLAME extractor record video from a webcam for a duration $T_{video}$, batch-processes, and then publish at $F_{fps}$.

The behavior generation module receives these multimodal data streams parallely to predict the next sequence of behavioral animation. LiveFLAME visualizer subscriber to this behavior predictor module directly as it can visualise predicted flame sequences. FastAPI backend  handling facial motion transformation and publish to the VuppetMaster receives data from the backend to animate the ARKit-based VIA. Throughout this system, each module maintains low latency in its data receiving data, processing, and publishing data. Additionally, we quantified the time taken by the interactive agents to animate 32 behavioral sequences, as depicted in Fig.~\ref{fig:realtime}(d).

\textbf{Performance Results:} Using a camera operating at both 30 fps and 24 fps, coupled with a four-microphone array processing 16-bit audio at 16 kHz, our system was configured for optimal performance. We set $T_{audio}=0.5$s for audio capture and subsequent Mel frequency processing. For video, $T_{video}=1$s was designated, allowing the FLAME extractor to batch-process and publish 30 or 24 images every second. The audio Mel frequencies were re-sampled at $4 \times F_{fps} \times T_{audio}$ fps, ensuring consistent feature processing. We have selected  $F_{fps}=30$ and $M_{fps}=120$, regardless of capturing image rate. The results are illustrated in Fig.~\ref{fig:realtime}.

\section{Discussion}
\label{section:Discussion}
We present the first framework that allows for real-time interaction with virtual avatars driven by deep learning-based behavior generation. As such, we address a crucial need that is pointed out in recent publications on neural behavior generation, namely the lack of a possibility to evaluate, develop, and apply such architectures in interactive human-agent scenarios \cite{ng2022learning2listen, Jonell2020}. 

\subsection{On Performance}

The results, as depicted in Fig.~\ref{fig:realtime}(a), elucidate the performance metrics of each module in our framework. The Behavior generator module showcased a swift processing time of 0.06 seconds, underscoring its real-time behavior generation efficiency. The Webcam FLAME extractor module registered batch-wise processing times of 0.79 and 0.83 seconds, contingent upon image quantity for 3D reconstruction, marking the highest computational latency. Still, this latency remains conducive for real-time applications as the input latency is higher than the output throughput. A noteworthy initial delay of 1s, highlighted in Fig.\ref{fig:realtime} (a), arises from batch processing of audio-video data. This latency is primarily attributed to $T_{video}$, suggesting potential reductions by optimizing its value, though it's computationally demanding. While the FLAME processing could match camera frame rates, real-time performance was not feasible on an i7 machine. Consequently, we opted for batch-wise processing at 1s intervals. Despite an initial 1 s delay, our approach reliably delivers outputs at either 30 fps or 24 fps.

The Mel frequency extractor demonstrated a swift processing time of 0.01 seconds, with a capture window rate of 0.00054 per frame, underscoring its real-time audio data processing efficiency as depicted in Fig.~\ref{fig:realtime}(a). The FastAPI server with $\mathcal{GL}$ transformation module clocked in at 0.0015 seconds for processing and 0.0019 seconds for publishing, highlighting its rapid data transformation and transmission capabilities. Fig.~\ref{fig:realtime} showcases the animation speeds for virtual agents, with real-time performance ranging between 20 ms to 50 ms. The animation frame rate is modifiable in our framework to achieve realistic transformed facial motions.

Interpreting these results necessitates an understanding of system-specific processing variability. Our framework emphasizes real-time suitability in interactive settings by ensuring low latency and effective data processing. The adjustable delays ($T_{video}$ and $T_{audio}$) derive from the method proposed in \cite{ng2022learning2listen}, which requires a 32 video frame sequence for output. Adapting to a 1s $T_{video}$ for cameras at 30 or 24 fps, we've modified the overlapping stride of FLAME and MFCC features in the behavior predictor. When the camera fps deviates from the trained 32-frame video sequence and $4 \times 32$ for MFCC as per \cite{ng2022learning2listen}, we incorporate features from preceding batches for the L2L prediction, ensuring modality compatibility. This adaptability lets our system handle varying fps while delivering uninterrupted output.


\subsection{Applications}
The proposed real-time framework offers a versatile solution for applications demanding interactive and immersive experiences. Potential use cases include virtual assistants, teleconferencing, educational and training platforms, and healthcare settings such as telemedicine, and virtual consultations. The framework's capacity to capture and represent users' facial expressions and speech enable the development of engaging, realistic interactive agents, fostering enhanced user experiences and facilitating more effective human-computer interactions across various domains. While our framework is agnostic to the concrete application scenario, we provide pre-trained models that can be valuable to users. We include a model trained on psychotherapy interactions as these interactions are (1) rich in social cues and interpersonal synchronization, and (2) difficult to obtain by most researchers due to data privacy considerations. We opted to train a model for a single specific therapist to represent nuanced individual behavior instead of a model interpolating between different persons. This is in line with the Learning to Listen approach presented by Ng and colleagues~\cite{ng2022learning2listen}, who trained individual models for TV presenters. ReNeLiB is designed for modularity, adaptability, and universality, catering to diverse virtual agents and platforms. While our behavior generation module utilizes the approach from~\cite{ng2022learning2listen}, its modular design ensures compatibility with alternative behavioral prediction techniques, such as~\cite{Jonell2020}.

\subsection{Limitations and Future Work}


The current framework exhibits certain limitations, such as the VuppetMaster character animation, which is managed by iteratively setting expression key values and head rotation values through web-based JavaScript. For more fluid movements, it would be advantageous to transmit a sequence of animations directly to the VuppetMaster engine, necessitating collaboration with the developers of VuppetMaster. Moreover, the system generates 32 frames of animation sequences per second in an autoregressive manner; however, it does not interpolate between predicted animation sequences to achieve smoother behavior, presenting an opportunity for enhancement.
In future work, the framework will undergo evaluation and user studies to assess the contextual appropriateness of generated behavior and to develop a standardized platform for evaluating listener behavior in interactive agents. These efforts will not only validate the framework's effectiveness but also contribute to its enhancement, allowing it to better accommodate a diverse array of intelligent IVA.

\begin{acks}
We would like to express our heartfelt gratitude to Prof. Dr. Cord Benecke and the diligent team at the University of Kassel, Germany, for their indispensable support. Their efforts, particularly in recording the therapy sessions, significantly helped our research. It is essential to note that the machine learning trained in our work do not retain any personal or health-related information. The anonymized data was utilized solely to derive facial behaviors, expressions, and raw audio features using MFCC. We strictly refrain from sharing any video or audio data with external entities and computation were conducted with Secure Machine Learning Architecture~\cite{semla:online} specialised for sensitive data processing. This research has been generously supported by the German Federal Ministry for Education and Research (BMBF) as a segment of the UBIDENZ project, under grant number 13GW0568D. Further, P. M\"uller's contributions were funded by the BMBF under grant number 01IS20075.
\end{acks}


\bibliographystyle{ACM-Reference-Format}
\balance
\bibliography{main}

\appendix

\section{Details of Real-life Therapy Dataset}\label{appendixA}

\begin{table*}[h!]
\caption{\textit{Total number of video hours per therapist sessions. \textbf{M} indicates male and \textbf{F} denotes female.}}
\label{tab:therapies}
\centering
\small 
\begin{tabular}{|c|c|c|c|c|c|}
\hline
\textbf{Therapist ID} & \textbf{No. Sessions} & \textbf{Ttl Duration(h)}& \textbf{Ttl Speech(h)} & \textbf{Patient(h)}& \textbf{Therapist(h)} \\ \hline
$TherapistA$ (M) & 49 (42F, 7M) & 75  & 54 & 36 & 18 \\ \hline
$TherapistB$ (F) & 51 (42F, 7M) & 70  & 53 & 43 & 10 \\ \hline
$TherapistC$ (F) & 22 (21F, 1M) & 22 & 17  & 14 & 3  \\ \hline
$TherapistD$ (M) & 12 (12F, 4M) & 14  & 12 & 10 & 2 \\ \hline
\end{tabular}
\end{table*}

This section provides a comprehensive description of the dataset utilized in this study, which was derived from an extended video corpus collected by \citet{opd,Peham2015}. The original research~\cite{opd} contains data from 80 women, including 16 healthy controls. In our extended version, a total of 139 video sessions were obtained, from which 134 were chosen for audio and video feature extraction. The remaining five sessions were excluded due to poor recording quality (audio or video). The selected sessions encompass 23 male patients and 113 female patients.

Table \ref{tab:therapies} presents a detailed breakdown of the dataset, comparing the four therapists who conducted the sessions in terms of the number of sessions, total duration, patient speaking duration, and therapist speaking duration. The segments where patients are actively speaking are defined as the region of interest (ROI) and represent the period during which the therapist is actively listening. By focusing on these ROIs, our study aims to analyze and model the active listening behavior exhibited by therapists during their interactions with patients.

\section{Data Segmentation and Definition of ROI}\label{appendixB}

To accurately dissect therapist-patient interactions, it was imperative to differentiate between speech and non-speech segments, while optimally segmenting the sessions.

We utilized the speaker diarization tool, pyannote-audio~\footnote{\url{https://github.com/pyannote/pyannote-audio}}, to segregate speakers, clustering speech segments by their duration into distinct intervals: \(S_{\text{no-speech}}\), \(S_{\text{backchanneling}}\), \(S_{\text{short-speech}}\), and \(S_{\text{long-speech}}\). 
The distinguished speech intervals were:

\begin{align*}
S_{\text{backchanneling}} & : [0.5-2] \text{s} \\
S_{\text{short-speech}} & : [2-3] \text{s} \\
S_{\text{long-speech}} & : >3 \text{s}
\end{align*}


To extract facial gestures, we employed the EMOCA method \cite{EMOCA:CVPR:2021}, which builds upon the FLAME 3DMM model \cite{3DMM}. This method estimates parameters like head-pose, expression, and head-shape. Integrating with mediapipe~\cite{mediapipe:online} enabled real-time face detection during inference, with identity-agnostic outputs achieved by omitting shape coefficients.



\end{document}